\documentclass[showpacs,twocolumn,prl,aps,color]{revtex4}
\usepackage{graphicx}
\begin{document}

\title{Topological Order in Gutzwiller Projected Wave Functions for Quantum Antiferromagnets}

\author{Tao Li}
\affiliation{Department of Physics, Renmin University of China,
Beijing, 100872, P.R.China}
\author{Hong-Yu Yang}
\affiliation{Center for Advanced study, Tsinghua University
,Beijing 100084, China}
\date{\today}

\begin{abstract}
We study the topological order in RVB state derived from Gutzwiller
projection of BCS-like mean field state. We propose to construct the
topological excitation on the projected RVB state through Gutzwiller
projection of mean field state with inserted $Z_{2}$ flux tube. We
prove that all projected RVB states derived from bipartite effective
theories, no matter the gauge structure in the mean field ansatz,
are positive definite in the sense of the Marshall sign rule, which
provides a universal origin for the absence of topological order in
such RVB state.
\end{abstract}


%75.10.Jm Quantized spin models
%75.50.Ee Antiferromagnets
%75.40.Mg Numerical simulation studies
\maketitle

It is widely believed that the strongly correlated systems may
exhibit exotic ground state structures and support exotic
excitations. The study of variational wave functions provide a
unique way to uncover such exoticness. In the two dimensional
quantum antiferromagnet(2DQAF), it is proposed that a featureless
quantum spin liquid state called resonating valence bond(RVB) state
may be realized\cite{anderson}. The RVB state is a coherent
superposition\ of spin singlet pairs and can be written as

\begin{equation}
 |\texttt{RVB}\rangle=\sum_{\text {dimer
 covering}}a(i_{1}j_{1};\ldots;i_{n}j_{n})\prod_{k=1}^{n}(i_{k}j_{k}),
\end{equation}
in which $(ij)=\frac{1}{\sqrt{2}}(i\uparrow j\downarrow-i\downarrow
j\uparrow)$ is a spin singlet pair(valence bond) between site $i$
and $j$. It is argued that the RVB state may support fractionalized
excitation of spin $\frac{1}{2}$ spinon\cite{zou,kivelson}.

A systematic way to generate the RVB state is through Gutzwiller
projection of BCS-like mean field state\cite{anderson}. Such kind of
projected RVB states are now widely used in the variational study of
undoped and doped quantum
antiferromagnet\cite{yokoyama,gros,leetk,sorella}. An important
issue on the projected RVB state is to characterize its structure
and understand its excitation in physical terms. This issue is
addressed at the effective theory level by Wen who introduced the
notion of quantum order\cite{wen2}. However, an understanding at the
wave function level is still lack and it is unclear to what extend
is the prediction of effective theory applicable to projected wave
functions. Fortunately, the projective method to generate the
variational ground state also provides a systematic way to generate
the low energy excitations. A variational wave function for low
energy excitation is simply generated by Gutzwiller projection of
mean field excited states\cite{anderson}. Such a procedure has been
followed by a number of authors to study quasiparticle
excitations\cite{sorella1,yunoki,gros1,leetk,randeria} and
topological excitations\cite{ivanov,arun} on projected RVB wave
functions. In this paper, we use such a projective construction to
study the topological excitation and topological order on the
projected RVB wave functions.

The concept of topological order is the key notion to describe the
structure of a RVB state and to understand the fractionalization of
excitation on it\cite{kivelson,read,wen1,senthil}. The topological
order can be defined as the rigidity of a system against topological
excitation(dubbed vison) in the bulk of the system. On a multiply
connected manifold, the topological order manifests itself as the
topological degeneracy of the ground state, in which case the ground
state has a number of locally similar but globally distinct partners
that differ by whether or not a vison threads each hole of the
manifold\cite{wen2,ivanov}.

These ideas can be illustrated by the simple example of quantum
dimer model\cite{kivelson,read,ralko}. In the quantum dimer
model(QDM), the Hilbert space factorizes into even and odd
topological sectors by the number of valence bonds that intersect a
cut line which starts at the center of vison and ends at the
boundary of the system(or infinity). A vison can be generated simply
by reversing the sign of the amplitudes of odd sector dimer
configurations\cite{read}. The vison defined in this fashion behaves
as a $\pi$ flux tube for an unpaired spin. Thus the topological
order is intimately related to the coherent motion of fractionalized
spin excitation in the RVB background.

At the Gaussian level, the projected RVB states are described by the
slave Boson effective theories\cite{lee,wen1,wen2}. According to the
slave Boson effective theory, a system with $Z_{2}$ gauge structure
in the mean field ansatz has topological order and supports
fractionalized excitations, while a $Z_{2}$ flux tube plays the role
of the vison\cite{senthil}.

In their pioneering work, Ivanov and Senthil conjectured that a
$Z_{2}$ gauge structure in the mean field ansatz is also essential
for the corresponding projected RVB state to show topological
order\cite{ivanov,arun,sorella2}. Based on the phenomenological
$Z_{2}$ gauge theory of the underdoped cuprates\cite{senthil}, they
proposed that a vison can be constructed by projecting BCS state
containing a superconducting vortex. They find numerically that RVB
states derived from certain effective theory with $U(1)$ gauge
structure, more specifically, the nearest-neighboring d-wave RVB
state(NND state), does not exhibit topological order. This is taken
as evidence for their conjecture, since the vison - vortex analog is
ill defined when the pairing term is gauged away in a $U(1)$
effective theory.

In this paper, we try to relate the topological order of projected
RVB state to their phase structure in the Ising basis. We find the
absence of topological order in the NND-type RVB states can be more
naturally attributed to the Marshall sign rule of the wave function
in the Ising basis, rather than the $U(1)$ gauge structure in the
mean field ansatz.

The Marshall sign rule is a well known property of the
antiferromagnetic Heisenberg model on bipartite
lattice\cite{marshall,weng}. According to the rule, the wave
function of the ground state is real  in the Ising basis and its
sign is given by $(-1)^{N_{\downarrow}^{even}}$, in which
$N_{\downarrow}^{even}$ is the number of down spins in the even
sublattice. In this paper, we prove that all RVB states derived from
bipartite slave Boson mean field states satisfy such a sign rule, no
matter what is the gauge structure of the mean field ansatz. At the
same time, we show a vison can be constructed by projecting a mean
field state containing a $Z_{2}$ flux tube, rather than a
superconducting vortex used by Ivanov and Senthil. Combining these
two points allow us to show that RVB states derived from all
bipartite mean field states, such as the NND state, do not support
topological order. Thus the Marshall sign rule provides an universal
origin for the absence of topological order for RVB state defined on
bipartite lattices\cite{note}, while a Gaussian level effective
theory fails for such a state.

In the slave Boson theory of RVB states\cite{lee,wen1,wen2},
Fermionic slave particles $f_{i\alpha}$ are introduced to represent
the $SU(2)$ spin variable as
$\mathbf{S}_{i}=\frac{1}{2}\sum_{\alpha\beta}f_{i\alpha}^{\dagger}\mathbf{\sigma}
_{\alpha\beta}f_{i\beta}$. These slave particles are subjected to
the constraint of $\sum_{\alpha}f_{i\alpha}^{\dagger}f_{i\alpha}=1$
to be a faithful representation of the spin algebra. At the saddle
point level, a RVB state is given by a BCS-like mean field ansatz
for the slave particles

\begin{equation}
\mathrm{H_{MF}}=\sum_{\langle
ij\rangle}\left(\psi_{i}^{\dagger}U_{ij}\psi_{j}+\mathrm{H.c.}\right)+\sum_{i}\lambda_{i}\left(f_{i\alpha}^{\dagger}f_{i\alpha}-1\right),
\end{equation}
in which $\psi_{i}=\left(\begin{array}{c}
  f_{i\uparrow} \\
  f_{i\downarrow}^{\dagger}
\end{array}\right)$. Here, $U_{ij}=\left(
  \begin{array}{cc}
    -\chi_{ij}^{\ast} & \Delta_{ij} \\
    \Delta_{ij}^{\ast} & \chi_{ij} \\
  \end{array}
\right)$ denote the mean field RVB order parameters, while
$\lambda_{i}$ are the Lagrangian multipliers introduced to enforce
the mean field constraint. The fluctuation of the RVB order
parameters are treated by effective gauge theories\cite{wen1}. When
the gauge symmetry of the mean field ansatz is broken to $Z_{2}$,
the effective theory is a $Z_{2}$ gauge theory\cite{wen1,senthil}.
Since the $Z_{2}$ gauge fluctuation is gapped at the Gaussian level,
it is believed that a $Z_{2}$ effective theory describes a phase
with truly fractionalized excitations.

The ground state of the mean field Hamiltonian has the form of
Cooper pair condensate\cite{himeda},
\begin{equation}
|\Psi\rangle=\exp\left(\sum_{ij}a_{ij}(f_{i\uparrow}^{\dagger}f_{j\downarrow}^{\dagger}
-f_{i\downarrow}^{\dagger}f_{j\uparrow}^{\dagger})\right)|0\rangle,
\end{equation}
in which $a_{ij}$ is the wave function of a Cooper in real space.
The RVB state is given by Gutzwiller projection of the mean field
ground state,
\begin{equation}
|\mathrm{RVB}\rangle=\mathrm{P_{G}}|\Psi\rangle=\mathrm{P_{G}}\left(\sum_{ij}a_{ij}(f_{i\uparrow}^{\dagger}f_{j\downarrow}^{\dagger}
-f_{i\downarrow}^{\dagger}f_{j\uparrow}^{\dagger})\right)^{\frac{N}{2}}|0\rangle,
\end{equation}
in which $\mathrm{P_{G}}=\prod_{i}(1-n_{i\uparrow}n_{i\downarrow})$
and $N$ is the number of lattice site.

The topological order can be checked by examining on a tours the
orthogonality of the RVB states that differs in the number of
trapped visons in the holes of the torus. Here, we propose a more
transparent way to construct the vison wave function. In the
effective theory context, a vison is nothing but a $Z_{2}$ gauge
flux tube. Hence, we propose to construct the vison wave function
through Gutzwiller projection of mean field state with an inserted
$Z_{2}$ gauge flux tube. The only effect of the inserted $Z_{2}$
flux tube is to reverse the sign of mean field RVB order parameters
$\chi_{ij}$ and $\Delta_{ij}$ on bonds that intersect the cut line
defining the vison, which amounts to changing the boundary condition
across the cut line from periodic to anti-periodic(or vice versa).
In Ivanov and Senthil's work, the role of a trapped superconducting
vortex is also to change the boundary condition across the cut line.
Thus our construction is equivalent to theirs but is applicable in
more general situations.

Now we show that the absence of topological order in the NND-type
RVB state can be attributed the Marshall sign rule. The NND theory
belongs to the general class of $U(1)$ bipartite effective theory,
which contains a $U(1)$ gauge structure in the mean field ansatz and
is bipartite in the sense that $\chi_{ij}$ and $\Delta_{ij}$ only
connect sites in opposite sublattices. Below we show that the RVB
states derived from such theories satisfy the Marshall sign rule.
Since a $U(1)$ bipartite theory is still $U(1)$ bipartite with the
insertion of a $Z_{2}$ flux tube, the RVB state with a trapped vison
also satisfy the Marshall sign rule and will not be orthogonal to
the RVB state with no trapped vison in general case\cite{note}. Thus
RVB states in this class are unlikely to exhibit topological order.

For convenience's sake, we work in the gauge with only hopping term,
\begin{equation}
\mathrm{H}_{\mathrm{MF}}=\sum_{\langle
ij\rangle,\sigma}\left(\chi_{ij}f_{i\sigma}^{\dagger}f_{j\sigma}+\mathrm{H.c.}\right).
\end{equation}
The mean field ground state of $\mathrm{H_{MF}}$ is given by
\begin{equation}
|\Psi\rangle=\prod_{\xi_{m}<0}f_{m\uparrow}^{\dagger}f_{m\downarrow}^{\dagger}|0\rangle
\end{equation}
in which $f_{m}^{\dagger}=\sum_{i}\varphi_{m}(i)f_{i}^{\dagger}$
denotes the eigenstate of $\mathrm{H_{MF}}$ of eigenvalue $\xi_{m}$.
For a bipartite Hamiltonian, the eigenvalues appear in plus-minus
pairs. The corresponding eigenfunctions are
\begin{eqnarray*}
\varphi_{m}(i)  =\varphi_{m}^{e}(i)+\varphi_{m}^{o}(i)\\
\varphi_{\bar{m}}^{{}}(i) =\varphi_{m}^{e}(i)-\varphi_{m}^{o}(i),
\end{eqnarray*}
in which $\varphi_{m}^{e}(i)$ and $\varphi_{m}^{o}(i)$ are the
components of the eigenfunctions in the subspace of even and odd
sublattices. From the orthonormality of $\varphi_{m}(i)$, it is easy
to show that $\varphi _{m}^{e}(i)$ and $\varphi_{m}^{o}(i)$ form
complete and orthonormal basis in their respective subspaces. Thus,
the eigenvectors with $\xi_{m}<0$ suffice to expand the subspaces of
each sublattices. This property of the bipartite Hamiltonian is the
key to establish the Marshall sign rule in the projected RVB states.

In the basis expanded by
$\prod_{k=1,\frac{N}{2}}f_{i_{k}\uparrow}^{\dagger}f_{j_{k}\downarrow}^{\dagger}\left\vert
0\right\rangle $, the amplitude of the above RVB state is given by
the following Slater determinant
\[
\psi=\left\vert
\begin{array}{ccc}
\varphi_{1}(i_{1}) & \cdots & \varphi_{1}(i_{\frac{N}{2}})\\
\vdots &  & \vdots\\
\varphi_{\frac{N}{2}}(i_{1}) & \cdots &
\varphi_{\frac{N}{2}}(i_{\frac{N}{2}})
\end{array}
\right\vert \left\vert
\begin{array}
[c]{ccc}%
\varphi_{1}(j_{1}) & \cdots & \varphi_{1}(j_{\frac{N}{2}})\\
\vdots &  & \vdots\\
\varphi_{\frac{N}{2}}(j_{1}) & \cdots &
\varphi_{\frac{N}{2}}(j_{\frac{N}{2}})
\end{array}
\right\vert .
\]
For convenience's sake, we define a reference spin configuration in
which all up(down) spins sit in the even(odd) sublattice. The
amplitude of this spin configuration is given by
\[
\psi_{ref}=\left\vert
\begin{array}{ccc}
\varphi_{1}^{e}(i_{1}) & \cdots & \varphi_{1}^{e}(i_{\frac{N}{2}})\\
\vdots &  & \vdots\\
\varphi_{\frac{N}{2}}^{e}(i_{1}) & \cdots & \varphi_{\frac{N}{2}}^{e}%
(i_{\frac{N}{2}})
\end{array}
\right\vert \left\vert
\begin{array}{ccc}
\varphi_{1}^{o}(j_{1}) & \cdots & \varphi_{1}^{o}(j_{\frac{N}{2}})\\
\vdots &  & \vdots\\
\varphi_{\frac{N}{2}}^{o}(j_{1}) & \cdots & \varphi_{\frac{N}{2}}^{o}%
(j_{\frac{N}{2}})
\end{array}
\right\vert .
\]
Since the row vector
$(\varphi_{m}^{e}(i_{1})\cdots\varphi_{m}^{e}(i_{\frac {N}{2}}))$
form complete and orthonormal basis in the subspace of even
sublattice, the column vector
$(\varphi_{1}^{e}(i)\cdots\varphi_{\frac{N}{2} }^{e}(i))^{T}$ also
form a complete and orthonormal basis. The same is true for the odd
sublattice.

Now consider a general spin configurations. With no loss of
generality, we consider the spin configuration derived from the
reference configuration through exchange of the first $k$ up spins
and the first $k$ down spins. One easily verifies that the inner
product of the amplitudes for thess two configurations,
$\psi_{k}^{\ast}\psi_{ref}$, is given by the square modulus of a
$k$-th order determinant

\begin{equation}
\psi_{k}^{\ast}\psi_{ref}=| \left\vert
\begin{array}{ccc}
s(j_{1}i_{1}) & \cdots & s(j_{1}i_{k})\\
\vdots &  & \vdots\\
s(j_{k}i_{1}) & \cdots & s(j_{k}i_{k})
\end{array}
\right\vert | ^{2},
\end{equation}
in which
$s(j_{l},i_{n})=\sum_{m=1,\frac{N}{2}}\varphi_{m}^{o\ast}(j_{l})\varphi_{m}^{e}(i_{n})$,
 $i_{n}$ and $j_{l}$ denote the coordinates of the exchanged $k$ up
spins and $k$ down spins.

Thus, the amplitude of the RVB state derived from Eq.(5) is positive
definite up to a global phase. Taking into account the sign change
due to Fermion exchange, we arrive at the conclusion that the
projected RVB states derived from $U(1)$ bipartite theories satisfy
the Marshall sign rule and thus are very unlikely to exhibit
topological order. However, it is not clear whether the $U(1)$ gauge
structure or the bipartite nature of the theory is responsible for
the absence of topological order. To clarify this point, we examine
the phase structure of projected RVB state derived from an arbitrary
bipartite effective theory.

The mean field ansatz for an arbitrary bipartite theory is given by
\begin{eqnarray}
   \mathrm{H_{MF}}&=&\sum_{\langle ij\rangle,\sigma}\left(\chi_{ij}f_{i\sigma}^{\dagger}f_{j\sigma}+H. C.\right) \nonumber  \\
     &&+\sum_{\langle ij\rangle}\left(\Delta_{ij}(f_{i\uparrow}^{\dagger}f_{j\downarrow}^{\dagger}
     +f_{j\uparrow}^{\dagger}f_{i\downarrow}^{\dagger})+H.
     C.\right),
\end{eqnarray}
in which $\chi_{ij}$ and $\Delta_{ij}$ connect sites on different
sublattices and are otherwise arbitrary(note other details of the
lattice, such as its dimension, spatial symmetry and topology, are
all irrelevant for our discussion). The Lagrangian multipliers are
set to zero since the theory is particle-hole symmetric. To
facilitate the proof, we make a particle-hole transformation on down
spins\cite{yokoyama},

\begin{equation}
    f_{i\downarrow}\longrightarrow\epsilon_{i}\tilde{f}_{i\downarrow}^{\dagger},
\end{equation}
in which $\epsilon_{i}=1$ for $i\in$ even sublattice and
$\epsilon_{i}=-1$ for $i\in$ odd sublattice. The transformed
Hamiltonian reads

\begin{eqnarray}
 \mathrm{\tilde{H}_{MF}}&=&\sum_{\langle ij\rangle}
  \left(  \chi_{ij}f_{i\uparrow}^{\dagger}f_{j\uparrow}+H.c.\right)  +
  \sum_{\langle ij\rangle}
   \left(  \chi_{ij}^{\ast}\tilde{f}_{i\downarrow}^{\dagger}\tilde{f}_{j\downarrow}+H.c.\right)  \nonumber  \\
  &&+\sum_{\langle ij\rangle}\left(
  \epsilon_{j}\Delta_{ij}(f_{i\uparrow}^{\dagger}\tilde{f}_{j\downarrow}^{{}}-f_{j\uparrow}^{\dagger}\tilde{f}_{i\downarrow}^{{}})+H.c.\right).
\end{eqnarray}
The eigenvector of this Hamiltonian is given by
$\gamma_{n}^{\dagger}=\sum_{i}(u_{n}(i)f_{i\uparrow}^{\dagger}+v_{n}(i)\tilde{f}_{i\downarrow}^{\dagger})$.
Since the Hamiltonian is bipartite, its eigenvalues appear in
plus-minus pairs. The corresponding eigenvectors are given by
\begin{equation}
\left(
\begin{array}
[c]{c}
u_{n}(i)\\
v_{n}(i)
\end{array}
\right)  ^{even}\pm\left(
\begin{array}
[c]{c}
u_{n}(i)\\
v_{n}(i)
\end{array}
\right)  ^{odd},
\end{equation}
in which
$\left(
\begin{array}
[c]{c}
u_{n}(i)\\
v_{n}(i)
\end{array}
\right)  ^{even}$ and $\left(
\begin{array}
[c]{c}
u_{n}(i)\\
v_{n}(i)
\end{array}
\right)  ^{odd}$
are the projections of the eigenvectors in the even
and odd sublattices. For a bipartite Hamiltonian, these sublattice
projections form complete and orthonormal basis in their respective
subspaces. At the same time, since the system is spin rotational
symmetric, the eigenvalues are two fold degenerate. Specifically, if
$(u_{m}(i),v_{m}(i))$ is an eigenvector of the Hamiltonian, then
$(v^{\ast}_{m}(i),-u^{\ast}_{m}(i))$ is a eigenvector of the
Hamiltonian with the same eigenvalue. Thus, the mean field ground
state of the Hamiltonian can be written as
\begin{eqnarray}
  |\Psi\rangle=\prod_{\xi_{m}<0}\gamma_{m}^{\dagger}|\tilde{0}\rangle
= \prod_{m=1}^{\frac{N}{2}'}\left(  u_{m}(i)f_{i\uparrow}^{\dagger}+v_{m}(i)\tilde{f}_{i\downarrow}^{\dagger}  \right)  \nonumber \\
\times
\left(v_{m}^{\ast}(i)f_{i\uparrow}^{\dagger}-u_{m}^{\ast}(i)\tilde{f}_{i\downarrow}^{\dagger}\right)\left\vert
\tilde{0}\right\rangle,
\end{eqnarray}
in which $\zeta_{m}$ denotes the eigenvalue, and $\left\vert \tilde
{0}\right\rangle $ denotes the vacuum of $f_{i\uparrow}$ and $\tilde
{f}_{i\downarrow}$. The prime on the product indicates that the
product is taken over the eigenvectors with negative eigenvalues.

We now project the wave function into the subspace of no double
occupancy. At half filling, the up spins and the holes of the down
spins should occupy the same set of lattice sites in the projected
wave function. Thus a general spin configuration can be expressed in
the Fock basis as
$\prod_{m=1,\frac{N}{2}}f_{i_{m}\uparrow}^{\dagger}
\prod_{m=1,\frac{N}{2}}\tilde{f}_{i_{m}\downarrow}^{\dagger}\left\vert
\tilde{0}\right\rangle $. The amplitude of the projected RVB state
for this configuration is given by the following Slater determinant

\begin{equation}
\psi=\left\vert
\begin{array}
[c]{cccccc}
u_{1}(i_{1}) & \cdots & u_{1}(i_{\frac{N}{2}}) &
v_{1}(i_{1}) & \cdots &
v_{1}(i_{\frac{N}{2}})\\
\vdots &  & \vdots & \vdots &  & \vdots\\
u_{\frac{N}{2}}(i_{1}) & \cdots & u_{\frac{N}{2}}(i_{\frac{N}{2}}) &
v_{\frac{N}{2}}(i_{1}) & \cdots & v_{\frac{N}{2}}(i_{\frac{N}{2}})\\
v_{1}^{\ast}(i_{1}) & \cdots & v_{1}^{\ast}(i_{\frac{N}{2}}) &
-u_{1}^{\ast
}(i_{1}) & \cdots & -u_{1}^{\ast}(i_{\frac{N}{2}})\\
\vdots &  & \vdots & \vdots &  & \vdots\\
v_{\frac{N}{2}}^{\ast}(i_{1}) & \cdots & v_{\frac{N}{2}}^{\ast}(i_{\frac{N}%
{2}}) & -u_{\frac{N}{2}}^{\ast}(i_{1}) & \cdots &
-u_{\frac{N}{2}}^{\ast }(i_{\frac{N}{2}})
\end{array}
\right\vert .
\end{equation}
Following essentially the same steps that we have detailed in the
$U(1)$ case, one can show that the column vectors,
$(u_{1}(i_{e})\cdots
u_{\frac{N}{2}}(i_{e}),v_{1}^{\ast}(i_{e})\cdots
v_{\frac{N}{2}}^{\ast}(i_{e}))^{T}$ and $(v_{1}(i_{e})\cdots
v_{\frac{N}{2}}(i_{e}),-u_{1}^{\ast}(i_{e})\cdots
-u_{\frac{N}{2}}^{\ast}(i_{e}))^{T}$ for $i_{e}\in$ even sublattice,
and $(u_{1}(i_{o})\cdots
u_{\frac{N}{2}}(i_{o}),v_{1}^{\ast}(i_{o})\cdots
v_{\frac{N}{2}}^{\ast}(i_{o}))^{T}$ and $(v_{1}(i_{o})\cdots
v_{\frac{N}{2}}(i_{o}),-u_{1}^{\ast}(i_{o})\cdots
-u_{\frac{N}{2}}^{\ast}(i_{o}))^{T}$ for $i_{o}\in$ odd sublattice,
form complete and orthonormal basis in their respective subspaces.
From this orthogonality, one can show that
$\psi_{k}^{\ast}\psi_{ref}$ is still given by Eq.(7), with the
matrix element now given by $s(j_{l},i_{n})= \sum_{m=1,\frac{N}{2}}
(u_{m}^{\ast}(j_{l})u_{m}(i_{n})+v_{m}^{\ast}(j_{l})v_{m}^{\ast}(i_{n}))$.
As before, $\psi_{ref}$ denotes the amplitude of the reference spin
configuration with all up spins occupying the even sublattice and
all down spins occupying the odd sublattice, while $\psi_{k}$
denotes the amplitude of a general spin configuration with $k$ up
spins occupying the odd sublattice.  $i_{n}$ and $j_{l}$ denote the
coordinates of the exchanged $k$ up spins and $k$ down spins.

Thus, the amplitude of the RVB state derived from Eq.(8) in the
basis expanded by
$\prod_{k=1,\frac{N}{2}}f_{i_{k}\uparrow}^{\dagger}
\tilde{f}_{i_{k}\downarrow}^{\dagger}\left\vert
\tilde{0}\right\rangle $ is also positive definite up to a global
phase. It is easy to check that the Marshall sign rule has been
built into this basis. Thus projected RVB state derived from an
arbitrary bipartite effective theory satisfies the Marshall sign
rule and are very unlikely to show topological order, no matter the
gauge structure of the mean field ansatz. Since the topological
order is directly responsible for the existence of fractionalized
excitations, our result also implies that the Marshall sign rule
provides a universal origin of confining force for fractionalized
excitations on bipartite lattice.

Our proof of the Marshall sign rule makes it clear that a $Z_{2}$
gauge structure in the mean field ansatz alone is not sufficient for
the derived RVB state to show topological order. To illustrate this
point, we have checked the topological order in the RVB state
derived from a $Z_{2}$ bipartite theory with the following mean
field ansatz on square lattice
\begin{eqnarray}
  U_{i,i+x}=-\tau_{3}+\tau_{1};U_{i,i+y}=-\tau_{3}-\tau_{1} \nonumber \\
  U_{i,i+3x}==-\tau_{3}+\tau_{2};U_{i,i+3y}=-\tau_{3}-\tau_{2},
\end{eqnarray}
in which $\tau_{1},\tau_{2}$ and $\tau_{3}$ denote the three Pauli
matrices. In Figure 1a, we plot the overlap between the state with
periodic-antiperiodic boundary condition and that with
antiperiodic-periodic boundary condition on a torus. The result
shows clearly the absence of topological degeneracy in such an RVB
state. On the other hand, since the definition of vison is now
independent of the gauge structure of the effective theory, neither
is a $Z_{2}$ gauge structure in the mean field ansatz necessary for
the projected RVB state to show topological order. To illustrate
this point, we have checked the topological degeneracy in the RVB
state derived from a $U(1)$ nonbipartite theory with the following
mean field ansatz
\begin{eqnarray}
U_{i,i+x}=-\tau_{3};U_{i,i+y}=-(-1)^{i_{x}}\tau_{3};\nonumber\\
U_{i,i+x+y}=-3(-1)^{i_{x}}\tau_{3}.
\end{eqnarray}
This ansatz describes a $U(1)$ RVB state on anisotropic triangular
lattice. The isotropic version of this ansatz is found to be a good
approximation for the ground state of the Heisenberg model on
triangular lattice\cite{yunoki1}. The overlap for this RVB state is
shown in Figure 1b. The result indicates that the RVB state derived
from such a $U(1)$ theory exhibits topological degeneracy. Thus the
gauge structure in the mean field ansatz seems to be not a good
teller for existence of the topological order in the projected RVB
state.

\begin{figure}[h!]
\includegraphics[width=9cm,angle=0]{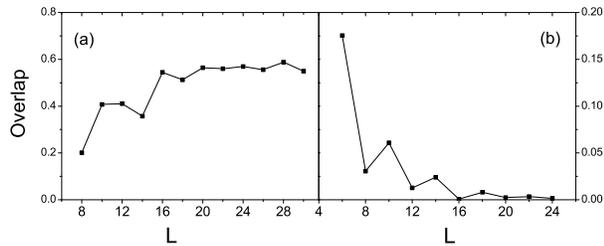}
\caption{Overlaps between the RVB state with periodic-antiperiodic
boundary condition and the RVB state with antiperiodic-periodic
boundary condition on a torus of size L.(a) $Z_{2}$ bipartite case,
(b)$U(1)$ non-bipartite case. }
 \label{fig1}
\end{figure}

In summary, we have proposed a new way to construct topological
excitation on projected RVB states. We find the Marshall sign rule
provides a universal origin for the absence of topological order for
RVB states derived from bipartite effective theories. This indicates
that a Gaussian level effective theory is insufficient for RVB
states defined on bipartite lattices.

This work is supported by NSFC Grant No.90303009. Discussions with
Z. Y. Weng, X. Q. Wang, L. Yu, M. Ogata and A. Paramekanti are
acknowledged.

\bigskip%

\bigskip


\begin{thebibliography}{99}
\bibitem{note} Our proof does not exclude the possibility that two
positive definite wave function be orthogonal in the thermodynamic
limit. Thus, the Marshall sign rule only provides a possible origin
for the absence of the topological order and is insufficient to
guarantee the absence of the topological order.
\end{thebibliography}
\end{document}